\documentclass[aps,prl,preprint,groupedaddress]{revtex4}
\begin{document}
%
%
%
%
\title{Detecting and Identifying Heavy Nuclei and Antinuclei with Standard Detectors}
\author{John Swain, Thomas Paul, Allan Widom}
\address{Department of Physics, Northeastern University, Boston, MA 02115, USA
}
\author{Y. N. Srivastava}
\address{INFN \& Dipartimento di Fisica, Universit\'{a} di Perugia, Perugia, Italy and\\
Department of Physics, Northeastern University, Boston, MA 02115, USA}
\date{Oct. 23, 2011}


\vspace*{-0.7cm}
\begin{abstract}
Most data gathered from high energy experiments at colliders
are analyzed assuming that particles
stable enough to not decay in the detector volume, and able to interact
strongly or electromagnetically, must be electrons, muons, protons,
neutrons, photons, kaons, and charged pions, or their antiparticles.
While light nuclei and antinuclei such as (anti)deuterons have been
detected,  we argue that it is experimentally interesting to look for
even heavier nuclei in high energy collisions.
To this end, we point out that using only tracking and calorimetry information
it is, in principle, possible to also search for high energy nuclei and antinuclei
and determine, with errors, their charge $Z$ and atomic weight $A$.
\end{abstract}
\maketitle

\section{Introduction}

Given the observation of antideuterons by the ARGUS\cite{argus1,argus2} and CLEO\cite{cleo}
collaborations in $e^+e^-$ collisions around center-of-mass energies around
10 GeV, the LEP collaborations OPAL\cite{opal} and ALEPH\cite{aleph} in $e^+e^-$
collisions around 90 GeV, as well as by the ZEUS\cite{zeus} and H1\cite{H1}
at HERA in DIS scattering events with center-of-mass energies around
300-318 GeV. 

The production of antideuterons is perhaps somewhat puzzling. Unlike deuterons,
they certainly cannot be produced by beam-gas or beam-wall collisions by simply
ejecting deuterons from nuclei. In whatever model one chooses for baryon
production it might seem somehow unlikely to produce 6 antiquarks close
enough in x-space to bind and form a proton and a neutron while not breaking up
any fragile deuteron produced -- its binding energy being a mere 2.2 MeV
which is far lower than the energies flying about in 10 GeV or 90 GeV $e^+e^-$
collisions!

In fact, the standard LUND fragmentation models \cite{LUND} while quite
successfully reproducing ratios of pseudoscalar and vector mesons from
string fragmentation ideas, require some degree of tuning to produce the correct
number of baryons such as protons and neutrons, let alone producing
the correct numbers of (anti)deuterons. There is a coalescence
model \cite{coalescence} developed for heavy ion collisions
which has been used
with some success in modelling (anti)deuteron production, but the point
of view we take here is that it would be both interesting and possible to 
look for much heavier (anti)nuclei. Certainly the fact that one sees
$\alpha$ decay from nuclei (and indeed this was one of the first nuclear
processes observed, with the $\alpha$ particle being much more stable
that a deuteron) is encouraging.

Higher baryon number bound states have also been produced. Notably,
the antihypertriton (an antineutron-antiproton-antilambda bound state) has
been seen by the STAR collaboration\cite{star} at RHIC in collisions
Au+Au collisions at $\sqrt{s}=200$GeV. The Phenix\cite{phenix} collaboration
at RHIC has also reported antideuteron production. ALICE has reported the production
of antideuterons, antitritons and antialpha particles in $pp$ collisions
at $\sqrt{s}=7$ TeV and in Pb-Pb collision at  $\sqrt{s}=2.76$ TeV.

It has been pointed out\cite{Baldini} that the $e^+e^-$ production of baryon-antibaryon
pairs, including $b\bar{p}$, $\Lambda_c\bar{\Lambda_c}$,
and $\Lambda_c\bar{\Lambda_c}$ at threshold
at BaBar \cite{Babar1,Babar2}
is consistent with a form factor of unity -- that is, despite being rather complex
objects made of many quarks and antiquarks, they are produced as if they
were point particles. This remarkable experimental fact might well give one
pause before dismissing complex heavy objects such as nuclei as being
intrinsically hard to make due to some putatively tiny form factor.

The suggestion has recently been made\cite{Higgs_to_nuclei} that there is
possibly interesting Higgs physics involving the production of heavy nuclei and
anti-nuclei with charges and masses far in excess of those carried by the usual
particles assumed to be produced.

Distinguishing an antideuteron from an antiproton is quite a challenging
task, but, as we shall see, heavier (anti)nuclei, if perhaps less likely
to be produced, may well be more amenable to experimental detection.

\section{Detecting Nuclei and Antinuclei}

We will assume a detector with tracking that involves
at least some determination of the charge deposited
per unit length, a magnetic field to allow the quotient
of momentum and charge to be determined. Measurements
of ionization give information on charge.
Calorimetry gives a measure of kinetic energy, and, possibly
in the case of annihilation of an antinucleus, total mass\cite{Grupen}.

Detecting and identifying a nucleus or antinucleus involves determining
its charge $Z$ and mass $M$. We briefly examine now how this might
be done.

\subsection{Determining $Z$}

The first simple observation is that for minimum ionizing particles
of charge $Z$, the most probable energy loss along a path is proportional
to $Z^2$. Responses from detectors typically have long tails
as represented by distributions such as the ones due to Landau
or Vavilov\cite{Grupen}, so there is always a possibility of misidentification of
$Z^2$, but for any given experimental setup it is possible
(though not often done) to construct a likelihood function
$L_{tracker}(Z)$ for the charge. Simple approximations can be
derived from a measured distribution for the signal read
out for a charge $Z$ based on the fact that the most likely
energy deposited scales as $Z^2$. 

Often detectors may be essentially blind to $Z^2$ by simply
not having the dynamic range to read out large signals which
would otherwise be taken to be fluctuations in the long tails
of energy loss distributions due to charge $|Z|=1$ particles, but
there is no problem in principle in determining the likelihood
$L_{tracker}(Z)$, and even allowing it to depend on the particle speed
$\beta$ to allow for the rapid increase in energy deposited at
lower speeds which scales as $1/\beta^2$. A determination
of $\beta$ is typically more difficult and involves additional
detector components such as time-of-flight systems or
Cerenkov or transition radiation detectors. Here we simply
note that if information on $\beta$ is available it can be
included into a likelihood function.

For this short note
we will restrict attention to highly relativistic nuclei and assume
that they are minimum-ionizing, but the general case is clearly
tractable. In the unhappy situation that the detector cannot 
determine energy deposited beyond some maximum corresponding
to the most likely, a likelihood function would still be constructible
but having little information about higher charge values.

The point is that it is, in principle, possible to determine the 
likelihood distribution of $|Z|$ for a particle such as a nucleus or
antinucleus,  in principle being strongly peaked at a given
value of $|Z|$.

The sign of $Z$ can be determined as is routinely done from
the curvature of the track that a particle makes in a magnetic field.

This means that $Z$ itself can be determined, or at least a likelihood
distribution for it obtained. Indeed, there is no intrinsic problem in
determining an inclusive ``charge spectrum'' for particles produced
in high energy collisions.

\subsection{Determining $M$}

The curvature $R$ of a track is related to its momentum $p_T$ transverse
to the magnetic field $B$ by 
\begin{equation}
[\frac{p_T}{GeV/c}] =.2998 [\frac{B}{Tesla}]  [\frac{R}{meters}].
\end{equation}
For a charge $Z$ measured in units of the
proton charge, this has to be corrected to 
\begin{equation}
[\frac{p_T}{GeV/c}] =.2998 [\frac{B}{Tesla}]  [\frac{R}{meters}]  [Z],
\end{equation}
 since tracks of particles carrying larger charges are bent more by a magnetic field.
The full momentum $\vec{p}$ is determined as usual from the direction
of the track and the transverse momentum.

The energy $E$ of a particle is given by $E^2=p^2+m^2$. For particles
of mass of a GeV/c$^2$ or less, and calorimetric errors typically greater than
a GeV/c$^2$, one can obtain little information about $M$.

For particles such as nuclei with masses much greater, the mass can
be determined by an apparent mismatch between energy calculated 
from $E\approx p$  and what is actually measured in the calorimeter.

The cases for nuclei and antinuclei are quite different. Let us consider
masses $M$ well over 1 GeV. For nuclei,
the rest mass of the nucleus will not be deposited as measurable
energy in the calorimeter, so one would find an energy seemingly
too {\it low} by $M$. For antinuclei, one expects complete annihilation
(mainly into $\pi^0$'s which decay into photons) and an apparent
{\em excess} in energy. 

For any specific experiment, a likelihood $L_{calorimeter}(Z,M)$ that the track was caused
by a nucleus or antinucleus of charge $Z$ and mass $M$
is clearly constructible. For a (anti)nucleus such as carbon-12, say, a
mismatch of 12 GeV could well be detectable with existing calorimeters.

\subsection{Putting it together}

Multiplying
likelihood distributions $L_{tracker}$ and $L_{calorimeter}$ gives a 2-dimensional
likelihood distribution $L_{total}(Z,M)$ , constructed using
existing techniques, for $Z$ and $M$ of a track being
due to a nucleus or antinucleus. 
Under the assumption of the particle indeed being a nucleus 
and (neglecting binding energies as small), the number of neutrons
can be approximated as $M-Z$ so a rather complete characterization
is possible. $L_{total}(Z,M)$ can be searched for peaks 
corresponding to nuclei or antinuclei either inclusively 
or in any given analysis. It can also be multiplied by whatever
prior distribution one favours and integrated in order to set
confidence levels on nuclei or antinuclei {\it not} being produced.

Of course in a practice one does not expect simple closed form
expressions for the relevant likelihoods. Generalizing the
usual analyses to allow for the possibility that tracks are produced
by heavy nuclei or antinuclei would be a major undertaking, but
one we feel would be an interesting and worthwhile one.

While the main point of this note is to show how nuclei and antinuclei
can be detected in high energy collider experiments, the ideas are
clearly applicable with little change to direct searches for fractionally
charged particles which may or may not decay in the calorimeter
system.

\section{Acknowledgements}

This work was supported in part by the US National Science Foundation
and grant NSF0855388.


\begin{thebibliography}{99}


\bibitem{argus1} H. Albrecht {\em et al.} (ARGUS Collab.), Phys. Lett. {\bf B157} (1985) 326.
\bibitem{argus2} H. Albrecht {\em et al.} (ARGUS Collab.), Phys. Lett. {\bf B236} (1990) 102.
\bibitem{cleo} D. M. Asner {\em et al.} (CLEO Collab.), Phys. Rev. {\bf D75} (2007) 012009. 
\bibitem{opal} R. Akers {\em et al.}  (OPAL Collab.) Z. Phys. {\bf C67} (1995) 203.
\bibitem{aleph} S. Schael {\em et al.} (ALEPH Collab.) Phys. Lett. {\bf B639} (2006) 192.



\bibitem{H1}  A. Aktas {\em et al.} (H1 Collab.) Eur. Phys. J. {\bf C36} (2004) 413.
\bibitem{zeus} S. Chekanov {\em et al.} (ZEUS Collab.) Nucl. Phys. {\bf B786} (2007) 181.

\bibitem{LUND} B. Andersson, ``The LUND Model'' (Cambridge University Press, 2005)
\bibitem{coalescence} S. T. Butler and C. A. Pearson, Phys. Rev. {\bf 129} (1963) 836.

\bibitem{star} B. I. Abelev {\em et al.}, Science, {\bf 328} (2010) 58.
\bibitem{phenix} R. Belmont (for the Phenix Collab.), Eur. J. Phys. {\bf C64} (2009) 243.

\bibitem{ALICE-antihypertriton} N. Sharma (for ALICE Collab.), arXiv:1109.4836 (Sept. 2011).

\bibitem{Baldini} R. Baldini {\em et al.} arXiv:0812.3283v2 [hep-ph].
\bibitem{Babar1} B. Aubert {\em et al.} (BABAR Collab.) Phys. Rev. {\bf D73} (2006) 012005.
\bibitem{Babar2} B. Aubert {\em et al.} (BABAR Collab.) Phys. Rev. {\bf D76} (2007) 092006.

\bibitem{Higgs_to_nuclei} in preparation.

\bibitem{Grupen} See, for example, C. Grupen and B. Shwartz,
Particle Detectors (Cambridge Monographs on Particle Physics, Nuclear Physics and Cosmology), 2011.



\end{thebibliography}
\end{document}